\documentclass[sigconf]{acmart}

\usepackage{subfigure}
\usepackage{booktabs} 
\usepackage[ruled]{algorithm2e}
\usepackage{graphicx}
\usepackage{multirow}
\usepackage{enumitem}

\usepackage{float}

\acmDOI{10.475/123_4}

\acmISBN{123-4567-24-567/08/06}

\acmConference[MM'17]{ACM international conference on Multimedia}{October 2017}
{Mountain View, CA USA}

\acmPrice{15.00}

\begin{document}
\bibliographystyle{phjcp}
\title{Hashtag-centric Immersive Search on Social Media}


%
\author{Yuqi Gao, Jitao Sang, Tongwei Ren, Changsheng Xu}
%
%
%


\begin{abstract}
Social media information distributes in different Online Social Networks (OSNs). This paper addresses the problem integrating the cross-OSN information to facilitate an immersive social media search experience. We exploit hashtag, which is widely used to annotate and organize multi-modal items in different OSNs, as the bridge for information aggregation and organization. A three-stage solution framework is proposed for hashtag representation, clustering and demonstration. Given an event query, the related items from three OSNs, Twitter, Flickr and YouTube, are organized in cluster-hashtag-item hierarchy for display. The effectiveness of the proposed solution is validated by qualitative and quantitative experiments on hundreds of trending event queries.
\end{abstract}

%
%


\maketitle

\section{Introduction}
Social media is recording and discussing what happens in real world. Its \emph{real-time information} and \emph{efficient propagation} has revolutionized the way people get access to their interested events, making various Online Social Networks (OSNs) the fundamental platform for information acquisition and sharing. In addition to real-time and propagation efficiency, the information on social media also features in its \emph{multi-source distribution}. Regarding the same event, relevant information distributes and propagates between different OSNs \cite{sang2016social}. For example,
regarding the event of 2016 US presidential election, people follow read-time progress on Twitter, watch and discuss debate video on Youtube, share inauguration photos on Instagram and Flickr.
These cross-OSN information enables comprehensive event description and understanding in different formats and from different perspectives.

In spite of the cross-OSN distribution characteristic, most social media search functions are single-OSN based and only support the exploration of information from one OSN. Taking Twitter search for example, although alternative search options are supported like time and popularity, the following issues prevent from a better experience. (1) Information richness. Popular OSN usually focuses on single modality, e.g., Twitter for text, Flickr for image, YouTube for video. It is reported from our data analysis that the images and videos embedded in Twitter tweets are not as good as those on Flickr and YouTube neither in quality nor in endorsement level. (2) Information coverage. OSNs describe information from different perspectives, which make complementary contribution to event understanding. While Twitter features in adequate data availability and the propagation efficiency, Flickr and YouTube haves advantage in information demonstration and social discussion, respectively.

\begin{figure}
\includegraphics[width=\linewidth]{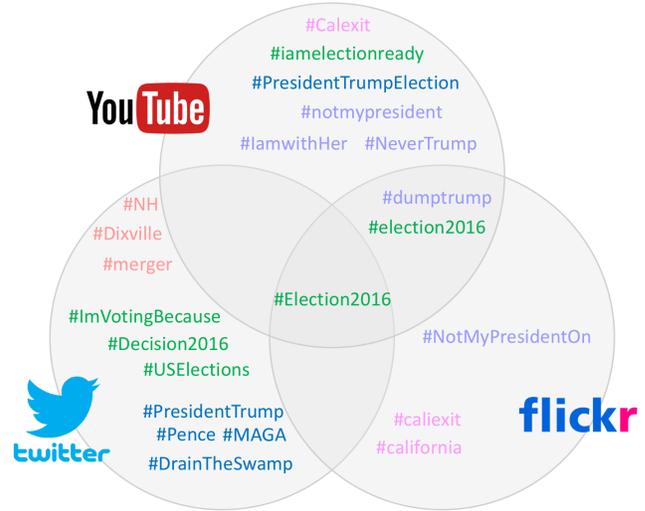}
\caption{The collected hashtags by issuing "Election 2016'' to different OSNs}
\label{QH}
\end{figure}

An immersive cross-OSN search framework is thus urgently needed: \emph{Given an event query, the related information from different OSNs are aggregated, organized, and demonstrated as search results.} The straightforward solution is to directly collect and organize the returned items from different OSNs. However, the processing at the individual item level suffers from several problems. (1) Relevance. It is difficult for common users to create an appropriate query to accurately describe the event. The items collected solely based on the relevance with an inaccurate query will make the search results noisy and biased. (2) Organization. OSNs support different search options and avoid an consistent solution.
Twitter supports searching by recentness and popularity, Flickr supports searching by date, interestingness and relevance, Youtube supports searching by date, relevance, rating and viewed times.
Moreover, the different modality focus compounds the difficulty in aggregation and organization.

\begin{figure*}
\includegraphics[width=\linewidth]{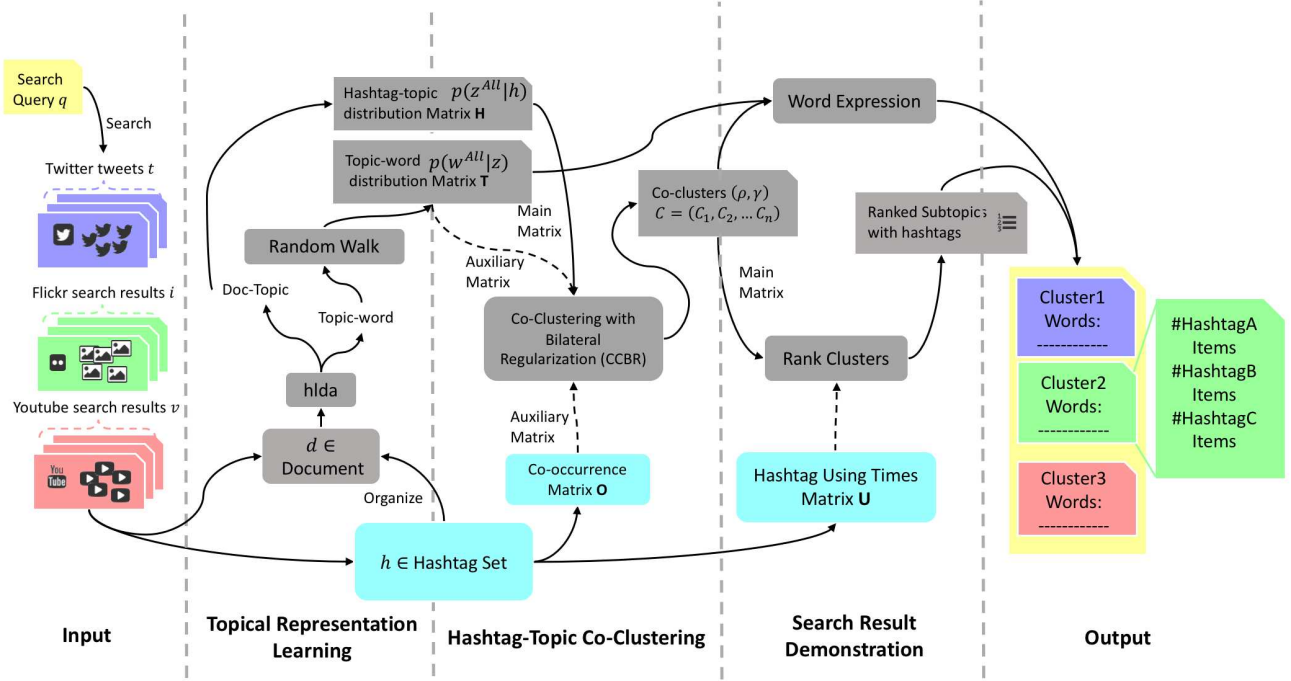}
\centering
\caption{The solution framework.}

\label{framework}
\end{figure*}

This paper proposes to exploit the hashtags as bridge to solve the cross-OSN immersive search problem. Hashtag is a typical social media feature widely used on different OSNs. It well addresses the above two issues regarding relevance and organization: (1) As a type of user annotation, hashtag guarantees the relevance of the annotated items to the events. Moreover, more related items can be retrieved by querying the hashtag. (2) Hashtag is originally adopted for information management, regardless of OSN or modality~\cite{yang2012we}, making it a natural tool for cross-OSN and multi-modal information organization. Fig.~\ref{QH} illustrates the collected hashtags from the returned search results of one query from Twitter, Flickr and YouTube where subtopics are marked with different colors. Two quick observations derive\footnote{We will justify the two observations with more evidence in the data analysis section.}: (1)Regarding the same event query, multiple hashtags are adopted on each OSN and vary between OSNs. (2)Different hashtags describe the different aspects, i.e., subtopics, of the event.

To consider the above observations and better exploit hashtag for cross-OSN information aggregation and organization,
the fundamental problem in our proposed solution is to discover the underlying subtopics, and organize the multiple hashtags as well as the hashtag-annotated items under the discovered subtopics. As shown in Fig.~\ref{framework}, the hashtag-centric immersive search framework consists of three stages.
The first stage learns topical representation over a unique vocabulary space for hashtags on each OSN. At the second stage, cross-OSN hashtags are clustered into subtopics considering both
the semantic correlation between topics and hashtag co-occurrence constrain. Finally, the derived hashtag clusters are ranked according to the relevance to the query for search result demonstration.  In the following we summarize the main contributions of this work:
\begin{itemize}[leftmargin=17pt]
\item We position the problem of cross-OSN immersive search. Information in multiple modalities and from different OSNs is integrated and demonstrated around event queries.
\item We propose a three-stage framework to exploit the hashtag as bridge for cross-OSN information integration and demonstration. The popularity and relevance warranty of hashtags enable an efficient and effective solution.
\item We implemented an online demo for search result demonstration\footnote{\url{https://hashtagasbridge.github.io/Hashtag/}}. Real-world quantitative and qualitative evaluation demonstrates the advantage of the proposed solution.
\end{itemize}

\section{Related work}
The topic of comparing and fusing search results from multiple search engines has been addressed in several studies. In~\cite{Jansen:2006:WSW:1138797.1138813}, the authors examined the characteristics of nine search engine logs in US and Europe. In~\cite{liu2009crowdreranking}, a crowd-ranking method is proposed to fuse the search results from different search engines for visual search re-ranking. Regarding the comparison between traditional Web search and social media search, \cite{teevan2011twittersearch} made a comparison of users search behaviors in Twitter search and Web search, and \cite{de2014seeking} examined the difference between traditional search engines and social media search in view of health related information. However, the topic of exploiting the search results from different social media networks has been largely ignored.

Cross-OSN analysis and application has recently received attention. One important research line is user-centric, i.e., to integrate the same individual's cross-OSN information for user modeling. The authors in \cite{Hsieh:2016:IRN:2872427.2883006} introduced an immersive cross-OSN solution to construct unified user profiles by associating user information on Facebook and Twitter. \cite{yang2015pinterest,yan2015unified} addressed the topic of cross-OSN recommendation by mining users' interests between Twitter, YouTube and Pinterest. The other research line, which is content-centric and more relevant to this work, is to connect the topic/event information across different OSNs. In~\cite{roy2012socialtransfer}, the authors proposed SocialTransfer, which is a cross-domain real-time learning framework to connect between Twitter and YouTube. A crowdsourcing solution is presented in \cite{yan2014mining} to discover the cross-OSN topic correlations. Inspired by these pilot studies, in this work, we propose to address the topic of cross-OSN search result fusion and demonstration.

Although hashtag was originally initiated by Twitter, it has become a common functionality across different OSNs. Many studies have analyzed hashtag usage pattern or employed hashtag for applications. \cite{yang2012we} examined the motivation, goal and usage patterns of users adopting hashtags. In~\cite{wang2014hashtag}, hashtag is analyzed and utilized for semantic organization and categorization. A recent work~\cite{Xing:2016:HSD:3016100.3016274} exploited hashtag to discover the fine-grained event-related semantics within Twitter.
These studies demonstrate the effectiveness of hashtag in social media information organization, which lay foundations and motivates us to implement a hashtag-centric solution in this work.

\section{Data Analysis}
To justify the cross-OSN search problem and motivate our solution using hashtag as bridge, this section conducts data analysis to first answer two questions: (1) What are the advantages of integrating single-OSN search results? (2) How people use hashtag across different OSNs?

\subsection{Single-OSN Search Comparison}
\begin{figure}[t]
\begin{minipage}[b]{0.49\linewidth}
\centering
\includegraphics[width=0.99\textwidth]{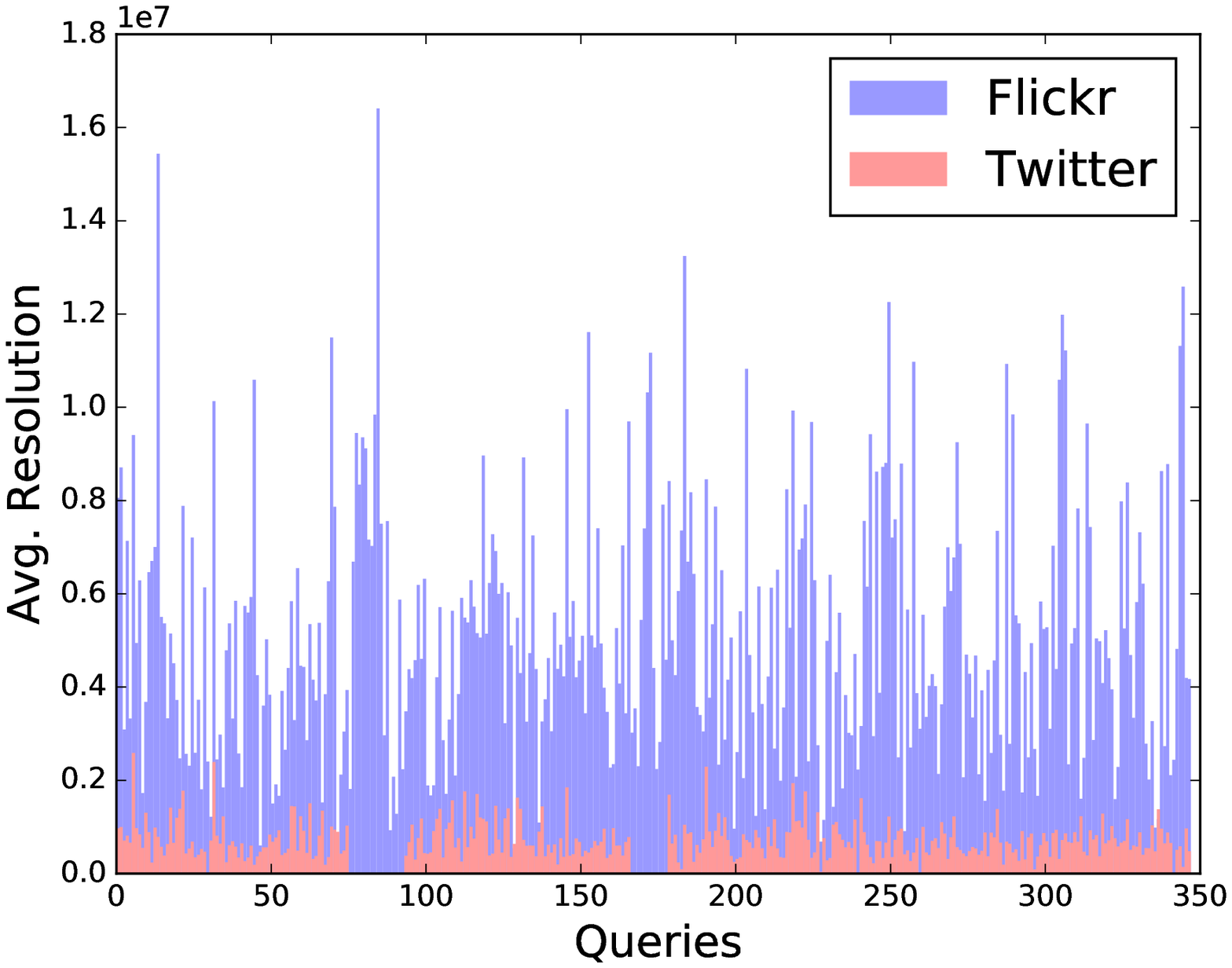}
\centerline{(a)Resolution of image}
\end{minipage}
\begin{minipage}[b]{0.49\linewidth}
\centering
\includegraphics[width=0.99\textwidth]{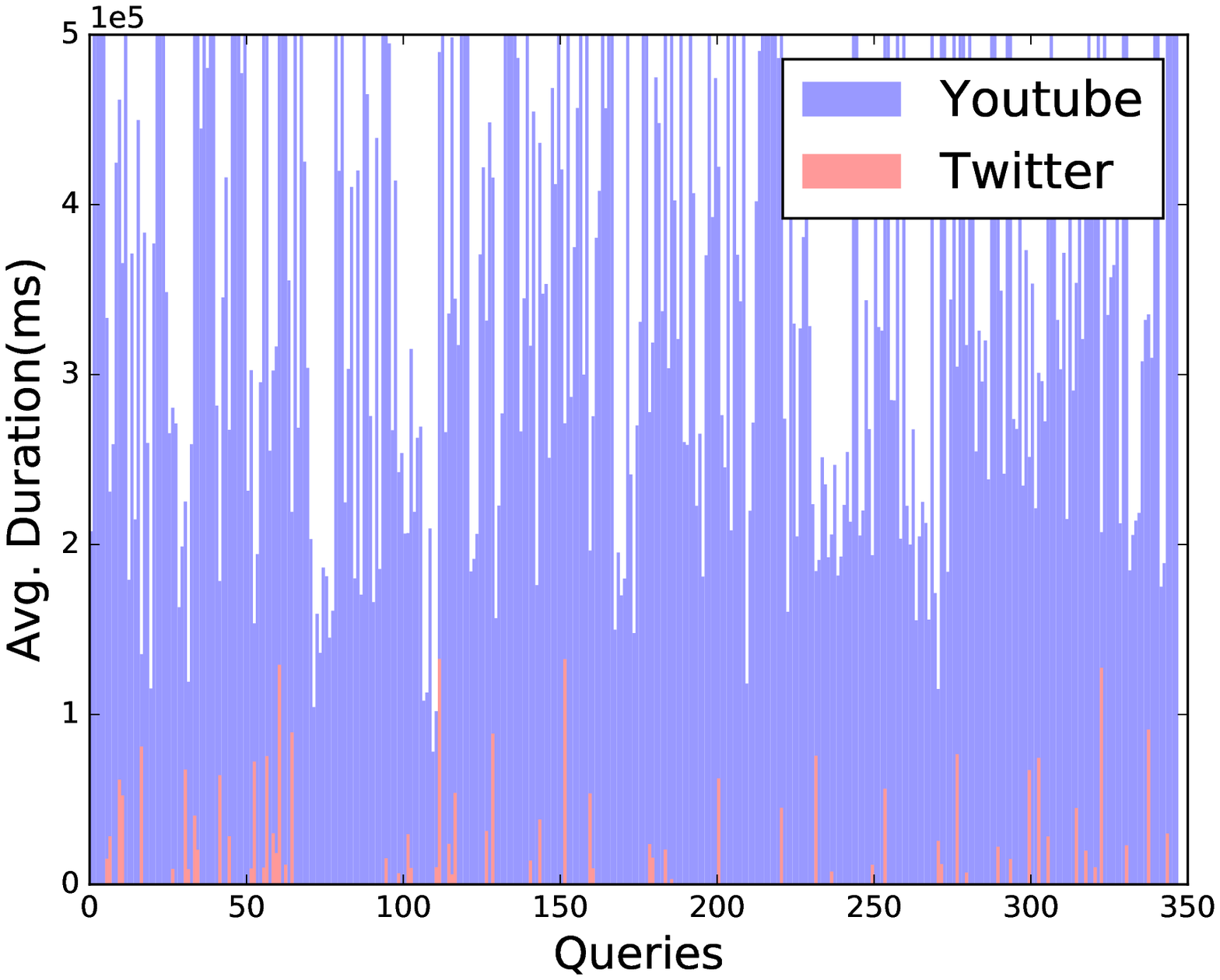}
\centerline{(b)Duration of video}
\end{minipage}
\caption{Information richness comparison.}
 \label{comparison}
 \end{figure}

We first examined the information richness of search results returned from different OSNs. Specifically, the resolution of shared images and duration of shared videos are compared among Twitter, Flickr and YouTube. 347 queries are selected from Google Trends\footnote{\url{https://trends.google.com}}, covering topics from politics, society, to economics and entertainment. Each query is submitted to the OSN APIs\footnote{
Flickr: \url{https://www.flickr.com/services/api/}\\
Twitter: \url{https://dev.twitter.com/overview/api}\\
Youtube: \url{https://www.youtube.com/yt/dev/api-resources.html}} to obtain totally 32,716 tweets from Twitter, 235,704 image items from Flickr and 195,505 video items from Youtube. Fig.~\ref{comparison}(a)(b) show the average resolution and duration of returned/embeded images and videos for each query, respectively. Regarding Twitter, only the tweets with embedded images or videos are counted when calculating average resolution/duration. It is easy to see that the search results from Flickr and YouTube capture significant richer information than those from Twitter in terms of image resolution and video duration. While Twitter has advantage in text-based information propagation, Flickr and YouTube can complement Twitter search results by providing more qualified images and videos.

\begin{figure}
\begin{minipage}[b]{0.49\linewidth}
\centering
\includegraphics[width=0.99\textwidth]{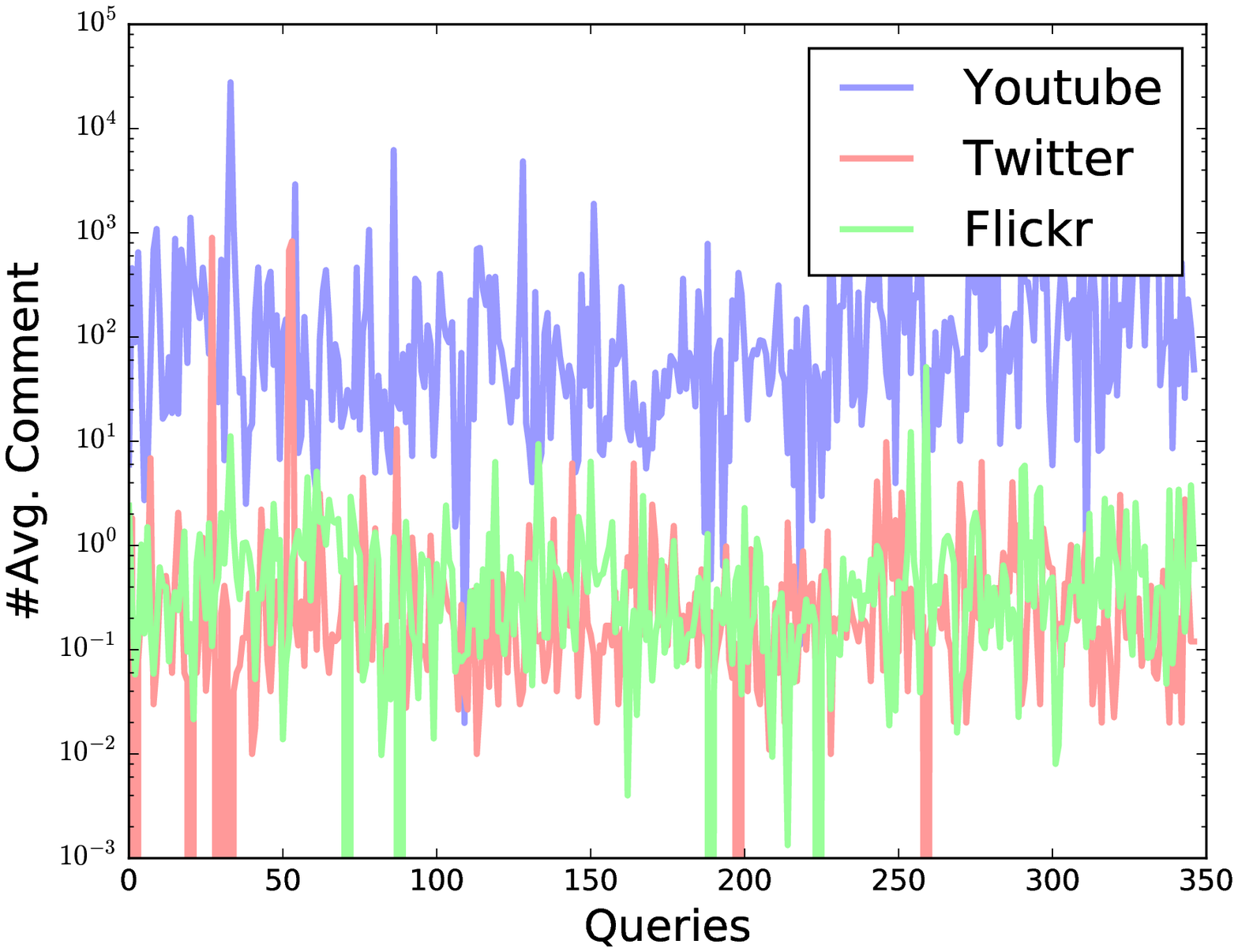}
\centerline{(a)Number of comments}
\end{minipage}
\begin{minipage}[b]{0.49\linewidth}
\centering
\includegraphics[width=0.99\textwidth]{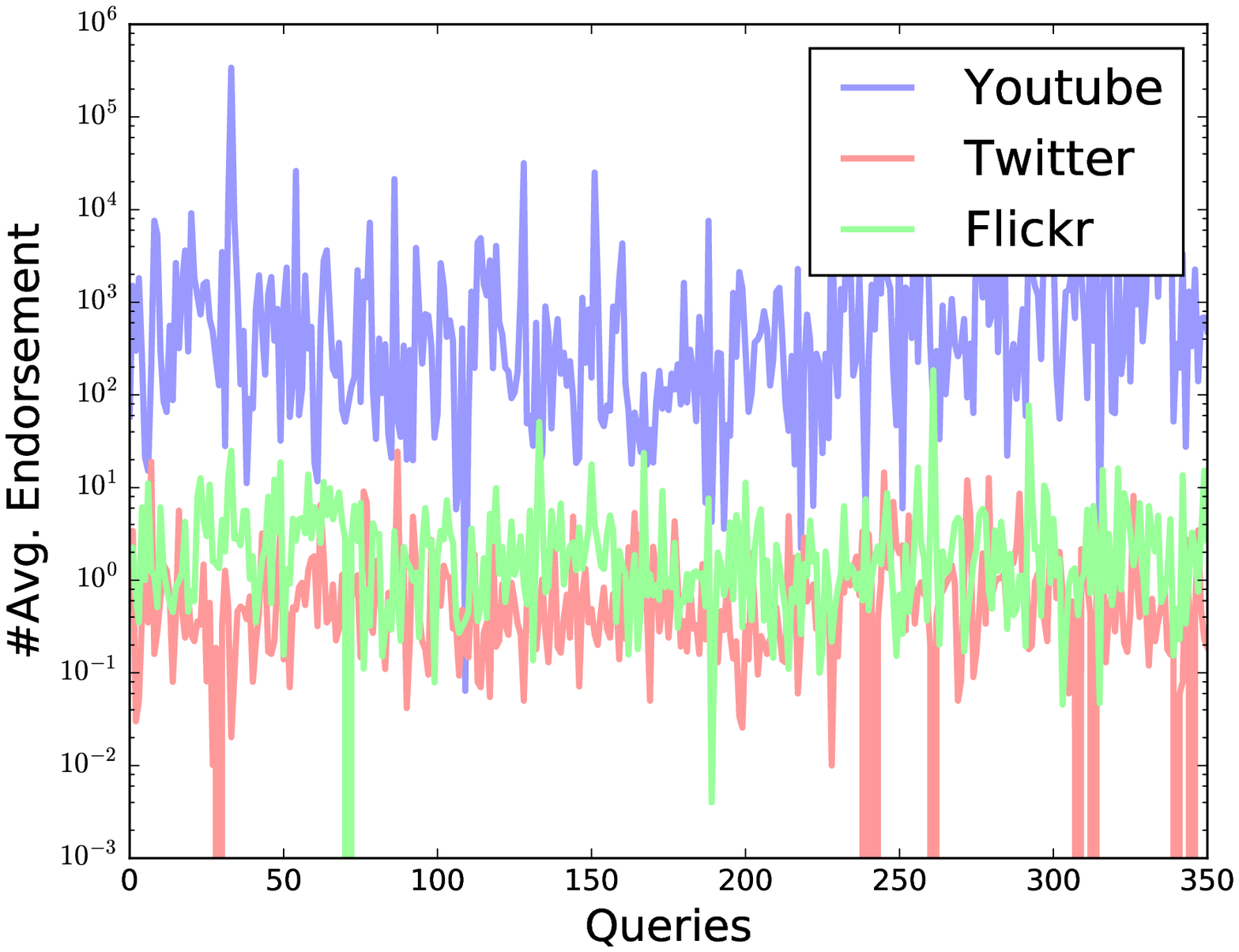}
\centerline{(b)Number of endorsements}
\end{minipage}
\caption{User interaction comparison.}
 \label{comparison2}
 \end{figure}

We then compared how different the search results of three OSNs attract user interactions. Two types of interactions are examined, i.e., comment and endorsement. For comment, the number of \emph{retweet} is calculated on Twitter. For endorsement, \emph{like}/\emph{dislike} are counted on YouTube, and \emph{favorite} is counted on Twitter and Flickr. Fig.~\ref{comparison2}(a)(b) illustrate the average number of comments and endorsements on the three OSNs in log-scale. We see that the two figures reach similar conclusions that YouTube search results generally attract more user interaction than Flickr and Twitter. Combining the above comparisons on information richness and user interaction likelihood, we justify the necessity of integrating single-OSN search results: not only a better multi-modal search experience is guaranteed, but more advanced features like social interaction can be explored and enabled.

\subsection{Cross-OSN Hashtag Usage Analysis}
\begin{figure}
\begin{minipage}[b]{0.49\linewidth}
\centering
\includegraphics[width=0.99\textwidth]{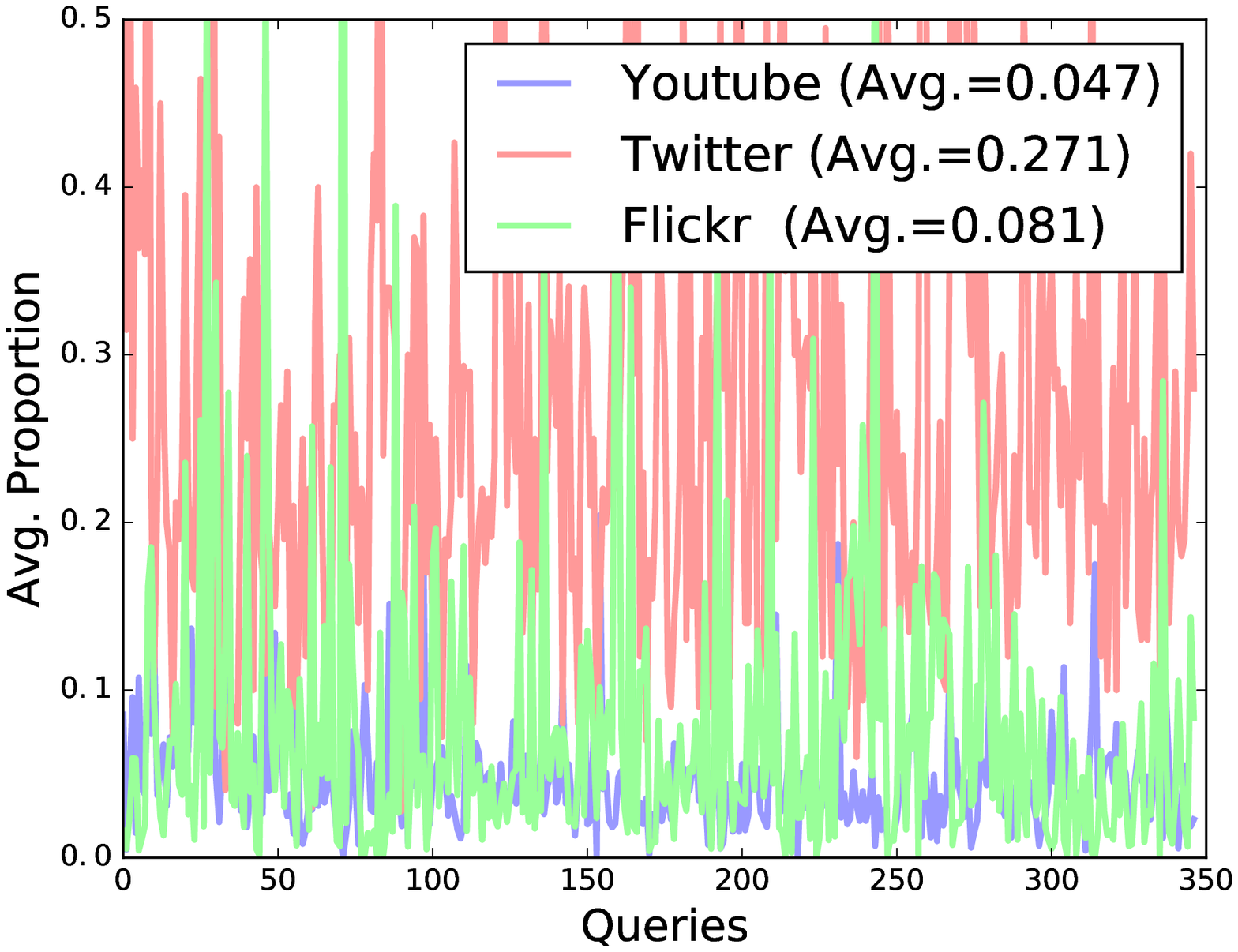}
\centerline{(a)Search results with hashtag}
\end{minipage}
\begin{minipage}[b]{0.49\linewidth}
\centering
\includegraphics[width=0.99\textwidth]{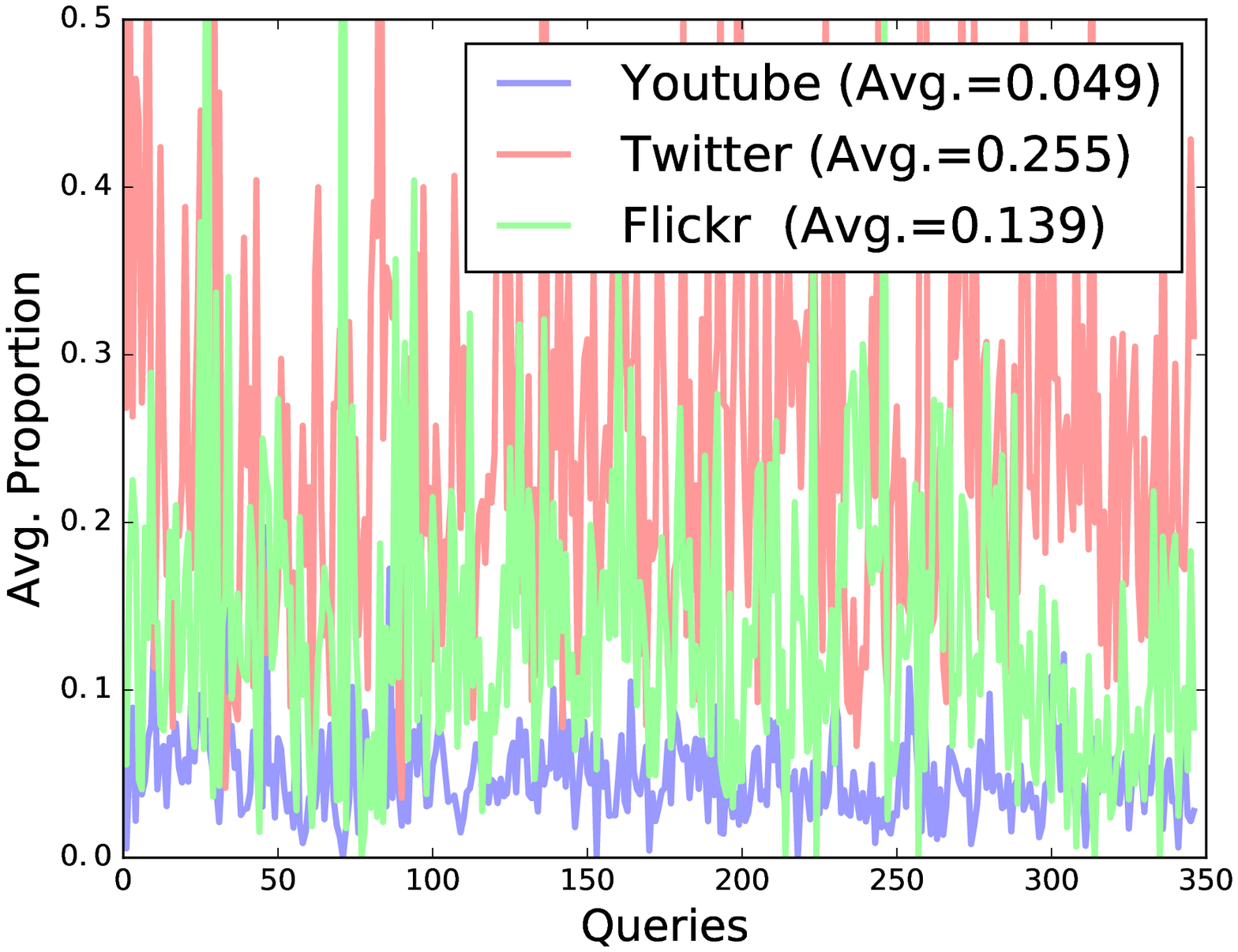}
\centerline{(b)User with hashtag}
\end{minipage}
\caption{Popularity of hashtag.}
 \label{comparison3}
 \end{figure}

This subsection addresses the availability and challenge of using hashtag to integrate cross-OSN search results. In Fig.~\ref{comparison3}, we calculate within the returned search results per query, what percentage of search results and users are with hashtag on the three OSNs. It is shown that hashtag is very popular on Twitter, with average search result and user percentage above 25\%. on Flickr, the percentage with hashtag varies between queries and the average percentage is 13.9\% for user and 8.1\% for search result. YouTube shows slightly lower hashtag popularity with about 4.7\% average percentage. Considering the relative importance of the hashtag-annotated search results\footnote{It is calculated that the $4.7\%$ YouTube videos with hashtag occupy over $10\%$ total endorsement. Moreover, in our solution, the items to be integrated are not limited to the API search results but inclusive of more items by retrieving the hashtags.}, this percentage is adequate for cross-OSN search result integration.

Other than popularity across different OSNs, hashtag also features in usage diversity. When talking about certain topics/events, users are likely to create multiple hashtags and these hashtags varies between OSNs. We counted the unique hashtags used per query on the three OSNs and summarize the number in Table~\ref{SF0}. It is noted that while multiple hashtags pose challenges to search result integration, they also provide possibility to fine-grained analysis by exploring subtopic from hashtags. We further compare the used multiple hashtags between OSNs. Spearman's footrule~\cite{diaconis1977spearman, bar2006methods} is widely used to measure the similarity of list pairs. We employed a normalized version calculated as follows:
\begin{equation}
NFr(\mu_1,\mu_2) = 1 - \frac{Fr^{|S|}(\mu_1,\mu_2)} {max\quad Fr^{|S|}}
\end{equation}
where $\mu_1,\mu_2$ are two lists, $|S|$ is the number of overlapping elements between two lists, $max\quad Fr^{|S|}$ equals $1/2|S|^2$ when $|S|$ is even and equals $1/2(|S|+1)(|S|-1)$ when $|S|$ is odd, $Fr^{|S|}(\mu_1,\mu_2)$ is the standard Spearman's footrule measure calculated as:
\begin{equation}
Fr^{|S|}(\mu_1,\mu_2) = \sum_{i=1}^{|S|}|\mu_1(i) - \mu_2(i)|
\end{equation}
where $\mu_1(i)$ is the rank of $i^{th}$ element in list $\mu_1$. The NFr score ranges from 0 to 1. The higher the NFr score, more similar between the two lists. To calculate the NFr score, we rank the unique hashtags returned from each OSN by the number of annotated search results in descending order. Table \ref{SF} shows the NFr score between OSNs averaged over the 347 queries on average. It is obvious that hashtag lists between OSNs are considerably different, making direct integration infeasible based on shared cross-OSN hashtags. Combining with the previous data analysis on popularity, we conclude that hashtag is widely used and available as bridge to integrate cross-OSN search results, but the integration and organization remain challenges due to the hashtag usage diversity.

\begin{table}
  \caption{\#. Unique hashtag used per query.}
  \label{SF0}
  \begin{tabular}{ccc}
    \toprule
     YouTube & Twitter &Flickr \\
    \midrule
   17.77 & 28.42 & 27.61\\
  \bottomrule
\end{tabular}
\end{table}

\begin{table}
  \caption{NFr score to examine hashtag usage difference between OSNs.}
  \label{SF}
  \begin{tabular}{ccc}
    \toprule
    Twitter\&Flickr & Twitter\&Youtube & Flickr\&Youtube\\
    \midrule
    0.1006 & 0.0857 & 0.0375 \\
  \bottomrule
\end{tabular}
\end{table}

\section{Solution}

\subsection{Topical Representation Learning}
In order to fully represent the hashtags as well as explore more related content, for each returned hashtag, we further collected all the items annotated by the corresponding hashtags (referred as \emph{extended search results}). The first stage learns hashtag topical representation from their annotated item set to facilitate the later hashtag clustering and ranking. Two issues are addressed: (1) Regarding the same query, most search results are related to the query and the returned hashtags may share a general topic from flat topic modeling method like Latent Dirichlet Allocation (LDA). We employ the hierarchical topic modeling method, hLDA~\cite{DBLP:conf/nips/BleiGJT03}, to explore fine-grained semantics and avoid the learned topical distributions of hashtags mixing with each other. ~(2) Topic modeling is separately conducted over the items on different OSNs. With different vocabularies, the learned topical distribution of cross-OSN hashtags cannot be directly compared. In this case, random walk is employed over the word semantic graph to bridge the cross-OSN topics with a unique integral vocabulary. In the following, we elaborate the solution to the above two issues respectively.

\subsubsection{Hierarchial Topic Modeling on respective OSN}~
\label{HierarchicalTopicModeling}
To facilitate the exploration of semantic structure, we conduct topic modeling over the extended search result set from each query $q$. The extended search results constitute three document collections $\mathcal{D}^T_q,\mathcal{D}^Y_q,\mathcal{D}^F_q$ for Twitter, YouTube and Flickr respectively. Hierarchical topic modeling is conducted over each OSN collection, with the textual content of each item $\mathbf{d}^{T,Y,F}_q$ as document\footnote{Textual content on each OSN is extracted as Twitter tweet, YouTube video title \& description and Flickr image title \& description.} over the respective vocabulary space $\mathcal{W}^{T,Y,F}$.

Different from the standard topic model like LDA which has a flat topic structure, the hierarchical topic model, i.e., hLDA, organizes topics in a tree of fixed depth $M$. Each document is assumed to be generated by topics on a single path from the root to a leaf through the tree. Note that all documents share the root topic in hierarchical topic model, which is consistent with the characteristics of search result collection. In our case, we select the tree depth $M=2$. After topic modeling, taking Twitter as example, each document $\mathbf{d}^T$ is attached with a 2-dimension topic distribution $[p(z^{T,root}|\mathbf{d}^T),p(z^{T,leaf}_k|\mathbf{d}^T)]$. $z^{T,root}$ is the root topic and $z^{T,leaf}_k$ is the $k^{th}$ leaf topic. For the $i^{th}$ hashtag $h^T_i$, its topical distribution over the Twitter leaf topic space is aggregated over all its annotated items and calculated as:
\begin{equation}
p(z^{T,leaf}_k|h^T_i) = \frac{\sum_{\mathbf{d}^T\in\mathcal{D}^T_{h^T_i}}p(z^{T,leaf}_k|\mathbf{d}^T)}{\sum_{k=1}^{K^T}\sum_{\mathbf{d}^T\in\mathcal{D}^T_{h^T_i}}p(z^{T,leaf}_k|\mathbf{d}^T)}
\end{equation}
where $K^T$ is the number of leaf topics on Twitter, $\mathcal{D}^T_{h^T_i}$ denotes the collection of items annotated with hashtag $h^T_i$. Noted that the root topics regarding the same query across OSNs are assumed to be similar, and we represent the hashtag by only considering the leaf topic distribution. As a result, we obtain three topic spaces $\{\mathbf{z}^{T,leaf}, \mathbf{z}^{Y,leaf}, \mathbf{z}^{F,leaf}\}$ over respective vocabulary set $\mathcal{W}^{T,Y,F}$ and each hashtag's distribution over the corresponding topic space.

\subsubsection{Random Walk-based Cross-OSN Vocabulary Integration}~
The goal is to analyze the cross-OSN topics over a unique vocabulary set $\mathcal{W}^{all}=\mathcal{W}^T\bigcup \mathcal{W}^Y\bigcup\mathcal{W}^F$. To bridge the different vocabulary sets, the semantic relation between words are considered. We make use of WordNet \cite{miller1995wordnet} to calculate the similarity $\pi_{ij}$ between word $w_i$ and $w_j$. With word $w\in\mathcal{W}^{all}$ as nodes and word similarity $\pi$ as weight, a word semantic graph $\mathbf{G}$ is constructed.

Random walk has been widely used in information retrieval~\cite{hsu2007video, liu2009tag, deng2012personalized} to explore the semantic correlations. In this work, we conduct random walk over the constructed word graph $\mathbf{G}$ to propagate the relevance scores among words. Specifically, a transition matrix $R_{|\mathcal{W}^{all}|\times |\mathcal{W}^{all}|}$ is constituted, where the transition probability from word $w_i$ to $w_j$ is calculated as $R_{ij} = \pi_{ij}/\sum_{w_k\in\mathcal{W}^{all}} \pi_{ik}$. At iteration $l$, the relevance score of node $i$ is denoted as $s_l(i)$, and the relevance scores of all word nodes constitute a vector $\mathbf{s}_l$ = $[...,s_l(i),...]^T$. The random walk process is thus formulated as:
\begin{equation}
\label{randomwalk}
\mathbf{s}_{l+1} = \alpha \sum_{i} \mathbf{s}_{l} R + (1-\alpha) \mathbf{t}
\end{equation}
where $\mathbf{t}$ denotes the initial probabilistic relevance scores as the original topic-word distribution, and $\alpha$ is a weight parameter that belongs to (0, 1).

The above process will promote the words with many close neighbors and weaken the isolated words. It is proved to converge to $\mathbf{s} = (1-\alpha)(\mathbf{l} - \alpha R)^{-1} \mathbf{t}$ which is a fixed point~\cite{liu2009tag}. After random walk, we obtain a cross-OSN topic space $\mathbf{z}^{all}$ over the integral vocabulary $\mathcal{W}^{all}$.

\subsection{Hashtag-Topic Co-Clustering}
This stage exploits the above-obtained topical distribution for cross-OSN hashtag clustering. Two issues remain: (1) Although topics from different OSNs are connected by integrating the vocabulary, each hashtag only has distribution over the topics on the corresponding OSN, e.g., for Twitter hashtag $h^T$: $p(\mathbf{z}^{F,leaf}|h^T)=\mathbf{0},p(\mathbf{z}^{Y,leaf}|h^T)=\mathbf{0}$. (2) Topics have intra-relation both within OSN and cross OSNs, which need to be considered for hashtag clustering. The intra-relation among topics are captured by the topic-word distribution over the unique vocabulary $\mathcal{W}^{all}$. To address the two issues, we introduce a hashtag-topic co-clustering solution, and incorporate topic semantic relation and hashtag co-occurrence information in topic and hashtag clustering respectively. The rest of the subsection first reviews the standard Bregman co-clustering method, and then elaborates our proposed hashtag-topic co-clustering solution.

\subsubsection{Bregman Co-Clustering}
Bregman co-clustering~\cite{banerjee2007generalized} is widely used in multi-dimensional data analysis. It aims to find the optimal row and column mapping $(\rho^{*},\gamma^{*})$ of an existing matrix $\mathbf{H}$ defined on two sets $\mathcal{H}$ and $\mathcal{T}$. Let $\nu=\{\nu_{ij};i=1,\cdots,|\mathcal{H}|,j=1,\cdots,|\mathcal{T}|\}$ be the joint probability measure of variable pair $(H,T)$ defining on $\mathcal{H}$ and $\mathcal{T}$ respectively, the element of matrix $\mathbf{H}$ takes values following $\nu$, i.e., $H_{ij}\sim \nu_{ij}$.

Let matrix $\hat{\mathbf{H}}$ be an approximation to $\mathbf{H}$, which depends only on mapping $(\rho,\gamma)$ and the resultant summary statistics such as row and column marginal, co-cluster marginals. Then the optimal co-clustering mapping $(\rho^{*},\gamma^{*})$ is identified such that the expected Bregman divergence with respect to $\nu$ between $\hat{\mathbf{H}}$ and $\mathbf{H}$ is minimized:
\begin{equation}
\label{coclustering}
\begin{split}
(\rho^{*},\gamma^{*}) &= \arg\mathop{\min}\limits_{\rho,\gamma} E[d_{\phi}(\mathbf{H},\hat{\mathbf{H}})] \\
& = \arg\mathop{\min}\limits_{\rho,\gamma} \sum_i\sum_j \nu_{ij} d_{\phi}(H_{ij},\hat{H}_{ij})
\end{split}
\end{equation}
where $\phi$ is a real-valued convex function and $d_\phi(z_1,z_2)$ is the \textit{Bregman divergence} defined as:
\begin{displaymath}
d_\phi(z_1,z_2) = \phi(z_1)-\phi(z_2)-<z_1-z_2,\bigtriangledown\phi(z_2)>
\end{displaymath}
$\bigtriangledown \phi$ is the gradient of $\phi$.

\subsubsection{Co-Clustering with Bilateral Regularization}
Bregman co-clustering can be iteratively solved. During each iteration, three subproblems are addressed. The first subproblem updates the approximation matrix $\hat{\mathbf{H}}$ by solving a \textit{Minimum Bregman Information} problem \cite{banerjee2007generalized} with current mapping $(\rho_i,\gamma_i)$. A permuted version of $\hat{\mathbf{H}}$ is then generated by randomly changing rows or columns, which is denoted as $\tilde{\mathbf{H}}$. The second and third sub-problems select the optimal column and row cluster mappings based on the generated permuted matrix $\tilde{\mathbf{H}}$. Specifically, the following two functions are optimized:
\begin{equation}
\label{coclusteringC}
\gamma_{i+1}(t) = \arg\mathop{\min}\limits_{1,...,L_{col}} E_{H|t}[d_{\phi}(\mathbf{H},\tilde{\mathbf{H}})]
\end{equation}
\begin{equation}
\label{coclusteringR}
\rho_{i+1}(h) = \arg\mathop{\min}\limits_{1,...,L_{row}} E_{T|h}[d_{\phi}(\mathbf{H},\tilde{\mathbf{H}})]
\end{equation}
where $L_{col},L_{row}$ denote the number of column and row clusters, $E_{H|t}, E_{T|h}$ denote the expectation according to marginal distribution of $\nu$ by fixing $T=t$ and $H=h$.

In the case of our problem, $\mathcal{H}$ and $\mathcal{T}$ represent the cross-OSN hashtag and topic sets. For each query, the matrix $\mathbf{H}_{N_{h}\times N_{t}}$ is constructed by the obtained hashtag-topic distribution $p(\mathbf{z}^{all}|\mathbf{h}^{T,Y,F})$, where $N_{h}, N_{t}$ denote the number of cross-OSN hashtags and topics. To conduct hashtag-topic co-clustering, we introduce a novel model of Co-Clustering with Bilateral Regularization (CCBR). At each iteration, the second and third sub-problems in Bregman co-clustering are modified by considering topic semantic relation and hashtag co-occurrence information, respectively.

The second sub-problem addresses column clustering, i.e., topic clustering. As in Eqn.~\eqref{coclusteringC}, the standard column clustering solution only exploits the involvement of different topics with hashtags recorded in $\mathbf{H}$. To incorporate the semantic relation between topics, we assume that the optimal topic cluster mapping $\gamma$ captures not only the topic-hashtag involvement but also the topic-word distribution. With the cross-OSN topic-word distribution $p(\mathcal{W}^{all}|\mathbf{z}^{all})$ constituting a matrix $\mathbf{T}_{N_{t}\times |\mathcal{W}^{all}|}$, we conduct row clustering over $\mathbf{T}$ simultaneously with the column clustering over $\mathbf{H}$. The updated optimal function is as follows:
\begin{equation}
\label{coclusteringC_new}
\gamma_{i+1}(t) = \arg\mathop{\min}\limits_{1,...,L_{col}} E_{H|t}[d_{\phi}(\mathbf{H},\tilde{\mathbf{H}})] + E_{W|t}[d_{\phi}(\mathbf{T},\tilde{\mathbf{T}})]
\end{equation}
where the second term on the right of the equation denotes the row clustering over $\mathbf{T}$, $E_{H|t}$ denote expectation according to marginal distribution by fixing $T=t$, and the generation of $\tilde{\mathbf{T}}$ is similar to that of $\tilde{\mathbf{H}}$. Note that since we are only interested to the row clustering of $\mathbf{T}$, the second term is actually modeled as one-sided Bregman clustering problem~\cite{banerjee2005clustering}, which is a special case of Bregman co-clustering by setting the number of column cluster as the same with the word number $|\mathcal{W}^{all}|$.

\begin{algorithm}[t]
\caption{Co-Clustering with Bilateral Regularization (CCBR)}
\label{BCAI}
\LinesNumbered
\KwIn{Hashtag-topic matrix $\mathbf{H}$, topic-word matrix $\mathbf{T}$, hashtag co-occurrence matrix $\mathbf{O}$, the number of column and row clusters $L_{col},L_{row}$.}
\KwOut{Optimal column and row mapping$(\mathbf{\gamma}^*,\mathbf{\rho}^*)$.}
Initialize $i=0$, randomly generate an initial $(\mathbf{\gamma}_i,\mathbf{\rho}_i)$;
\While{not converge \& not reach maximum iteration}
{
\textbf{Step A:} Update Minimum Bregman Information solution $\hat{H}$ with 	                   current $(\rho_t,\gamma_t)$\; Construct permuted $\tilde{H}$ for row and column respectively\;
\textbf{Step B:} Update topic cluster mappings $\mathbf{\gamma}$\;
\For{$t = 1$ to $N_t$}{
$\gamma_{i+1}(t) = \arg\mathop{\min}\limits_{1,...,L_{col}} E_{H|t}[d_{\phi}(\mathbf{H},\tilde{\mathbf{H}})] + E_{W|t}[d_{\phi}(\mathbf{T},\tilde{\mathbf{T}})]$
}
\textbf{Step C:} Update hashtag cluster mappings $\mathbf{\rho}$\;
\For{$h = 1$ to $N_h$}{
$\rho_{i+1}(h) = \arg\mathop{\min}\limits_{1,...,L_{row}} E_{T|h}[d_{\phi}(\mathbf{H},\tilde{\mathbf{H}})]+E[d_{\phi}(\mathbf{O},\tilde{\mathbf{O}})]$
}
$i\leftarrow i+1$.
}
$\mathbf{\gamma}^*=\mathbf{\gamma}_{i-1},\mathbf{\rho}^*=\mathbf{\rho}_{i-1}$.
\end{algorithm}

As for the row clustering sub-problem, i.e., hashtag clustering, we make assumption that the hashtags co-occuring in the same item have very high probability to belong to the same subtopic. To achieve this goal, we build a matrix $\mathbf{O}_{N_{h}\times N_{h}}$ with element $O_{ij}$ denoting the times that the $i^{th}$ and $j^{th}$ hashtags co-occur in the same item. To construct a unified formulation, similar to Eqn.~\eqref{coclusteringC_new}, we constrain that the optimal hashtag cluster mapping $\rho$ is also consistent with the clustering over $\mathbf{O}$. Therefore, the updated optimal function is as follows:
\begin{equation}
\label{coclusteringR_new}
\rho_{i+1}(h) = \arg\mathop{\min}\limits_{1,...,L_{row}} E_{T|h}[d_{\phi}(\mathbf{H},\tilde{\mathbf{H}})]+E[d_{\phi}(\mathbf{O},\tilde{\mathbf{O}})]
\end{equation}
By replacing Eqn.~\eqref{coclusteringC}\eqref{coclusteringR} with Eqn.~\eqref{coclusteringC_new}\eqref{coclusteringR_new}, the overall pseudo code of the proposed hashtag-topic co-clustering model is summarized in Algorithm~\ref{BCAI}.

\subsection{Search Result Demonstration}
After hashtag-topic clustering, for each query, we obtain $L_{row}$ hashtag clusters $\{C_{1},C_{2},...,C_{L_{row}}\}$, with each cluster consisting of $N_{C_l}$ hashtags from different OSNs. For each hashtag, a cluster-hashtag weight $p(h|C_l)$ is also derived to reflect the importance of hashtag $h$ to cluster $C_l$.

Demonstration consists of search result organization and search result description. Based on the hashtag clustering results, we propose to organize the search results following a cluster-hashtag-item hierarchy (illustrated in Fig.~\ref{demo}). Within hashtag, the items are organized chronologically. Within cluster, hashtags are ranked directly by the cluster-hashtag weight $p(h|C_l)$. For the ranking between hashtag clusters, their importance to describe the query is evaluated. Two factors are considered to calculate the cluster importance score. (1) The frequency that hashtags appear in the search result collection. Since the search results are basically relevant to the query, it is reasonable that the cluster with more hashtag-annotated items in the search result collection should be ranked higher. (2) The semantic relation between clusters. It is assumed that two hashtag clusters with similar semantic relation deserve close rankings to the query.

We first introduce how to evaluate the semantic relation between clusters. Given the cluster-hashtag weight $p(h|C_l)$ and the hashtag-topic distribution $p(z^t|h)$, it is easy to obtain the cluster-topic distribution $p(z^t;z^t\in \mathbf{z}^{all}|C_l)$ as:
\begin{equation}
p(z^t|C_l) = \sum_{h\in C_l} p(h|C_l)\cdot p(z^t|h)
\end{equation}
The semantic relevance $\kappa_{ij}$ between clusters $C_i$ and $C_j$ is then calculated as:
\begin{equation}
\label{Wij}
\kappa_{ij} = exp(-\frac{\sum_{z^t\in \mathbf{z}^{all}}(p(z^t|C_i)-p(z^t|C_j))^2}{2\sigma^2})
\end{equation}
where $\sigma$ is set to be the mean value of pairwise Euclidean distance between clusters.

The importance score of each cluster $\mathbf{\eta} = [\eta_1,\eta_2,...,\eta_{L_{row}}]$ is then obtained by minimizing the following cost function:
\begin{equation}
\label{Qcost}
Q(\mathbf{\eta}) = \sum_{i=1,j=1}^{L_{row}} \kappa_{ij} (\frac{1}{\sqrt{D_{ii}}} {\eta_i} - \frac{1}{\sqrt{D_{jj}}} {\eta_j})^2 + \psi \sum_{i=1}^{L_{row}}(\eta_i-U_i)^2
\end{equation}
where $D_{ii} = \sum_{j=1}^{L_{row}} \kappa_{ij}$, $\psi$ is weight parameter, and $U_i$ records the times that hashtags within cluster $C_i$ appear in the search result collection. The above problem can be solved by updating the importance score at each iteration\footnote{Detail for derivation is available in~\cite{lu2016tag}}:
\begin{equation}
\label{Rupdate}
\mathop{\mathbf{\eta}^{(t+1)}} = \frac{1}{1+ \psi} (\mathbf{\eta}^{(t)} S+\psi U)
\end{equation}
where $S = D^{-1/2}WD^{-1/2}$, $D = Diag(D_{11},D_{22},...,{D_{L_{row}L_{row}}})$, and $U = [U_{1},U_{2},...,U_{L_{row}}]$. After convergence, the derived $\mathbf{\eta}^{*}$ is used to rank the hashtag clusters for organization. Noted that the above cluster ranking strategy makes the noisy hashtag clusters rank very low due to their low appearance frequency and irrelevance with other clusters.

For search result description, since the hashtag clusters are expected to correspond to subtopics, we are motivated to generate semantic description for each hashtag cluster. Specifically, we calculate the cluster-word semantic distribution $p(w|C_l)$ by aggregating the cluster-topic distribution and topic-word distribution:
\begin{equation}
\label{clusterword}
p(w|C_l) = \sum_{z^t\in \mathbf{z}^{all}} p(z^t|C_l) \cdot p(w|z^t)
\end{equation}
As a result, each hashtag cluster (subtopic) can be represented by the 5-10 words with the highest $p(w|C_l)$.

\section{Experiments}
Based on the same dataset used for data analysis in Section 3, we reported in the following the experimental results of the three stages, respectively.
\subsection{Results of Representation Learning}
\begin{figure}
\includegraphics[width=\linewidth]{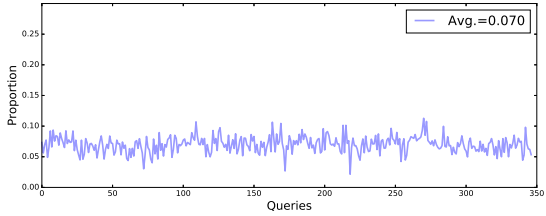}
\caption{Vocabulary overlap proportion.}
\label{topicmodelingfeatures}
\end{figure}

As introduced in the solution, given query $q$, hLDA is conducted on three document collections $\mathcal{D}^T_q,\mathcal{D}^Y_q,\mathcal{D}^F_q$ over the vocabulary space $\mathcal{W}^{T,Y,F}$ respectively. Empirical setting is used with $\alpha = 10$ $\gamma = 1$ and $\eta = 0.1$. After hLDA, random walk is conducted with $\alpha = 0.5$ to connect the respective vocabulary spaces to construct a unified vocabulary $\mathcal{W}^{all}$. Fig.~\ref{topicmodelingfeatures} shows the proportion of overlapped vocabulary $\mathcal{W}^{overlap} = \mathcal{W}^T \bigcap \mathcal{W}^Y \bigcap \mathcal{W}^F$ to $\mathcal{W}^{all}$. It is shown only about $7\%$ vocabulary is shared between the three OSNs, which validate the necessity for random walk-based vocabulary integration.

Table \ref{visual} illustrates some of the discovered leaf topics on different OSNs for the query ``Election 2016''. Each topic is represented by the top probable words, where the words appeared in the original vocabulary space is marked with black and the words extended by random walk from Twitter are highlighted with cyan, Flickr with blue and Youtube with red. Two observations derive: (1) The discovered topics have a wide coverage and the topics on different OSNs have some words and themes in common. (2) Random walk connects between different vocabulary spaces and enhances the topic representation with cross-OSN words.

 \begin{center}
 \begin{table}

\begin{tabular}{|c|c|}
 \hline
Platform & Topic  \\
 \hline
\multirow{2}{*}{Twitter}
                & ranks,states,worldpolitics,{\color{blue}meet}
                ...\\
                \cline{2-2}
                & published,{\color{red}newsletter},
                \color{black}{history,online,}...\\
\hline
\multirow{2}{*}{Flickr}
                & million,trump,election,votes,{\color{cyan}riches}...\\
                \cline{2-2}
                & nation,people,{\color{red}language},{\color{cyan}americanelection}...\\
\hline
\multirow{2}{*}{Youtube}
                & trump,donald,live,rally,{\color{cyan}presidenttrump},...\\
                \cline{2-2}
                & country,{\color{blue}children},
                \color{black}{feel,hold,}
                {\color{blue}citizens},...\\
                \hline
 \end{tabular}
 \vspace{+0.5cm}
 \caption{Visualization of topics from different OSNs of query "Election 2016".
 }
  \label{visual}
 \end{table}
 \end{center}

\begin{figure}
\begin{minipage}[b]{0.49\linewidth}
\centering
\includegraphics[width=0.99\textwidth]{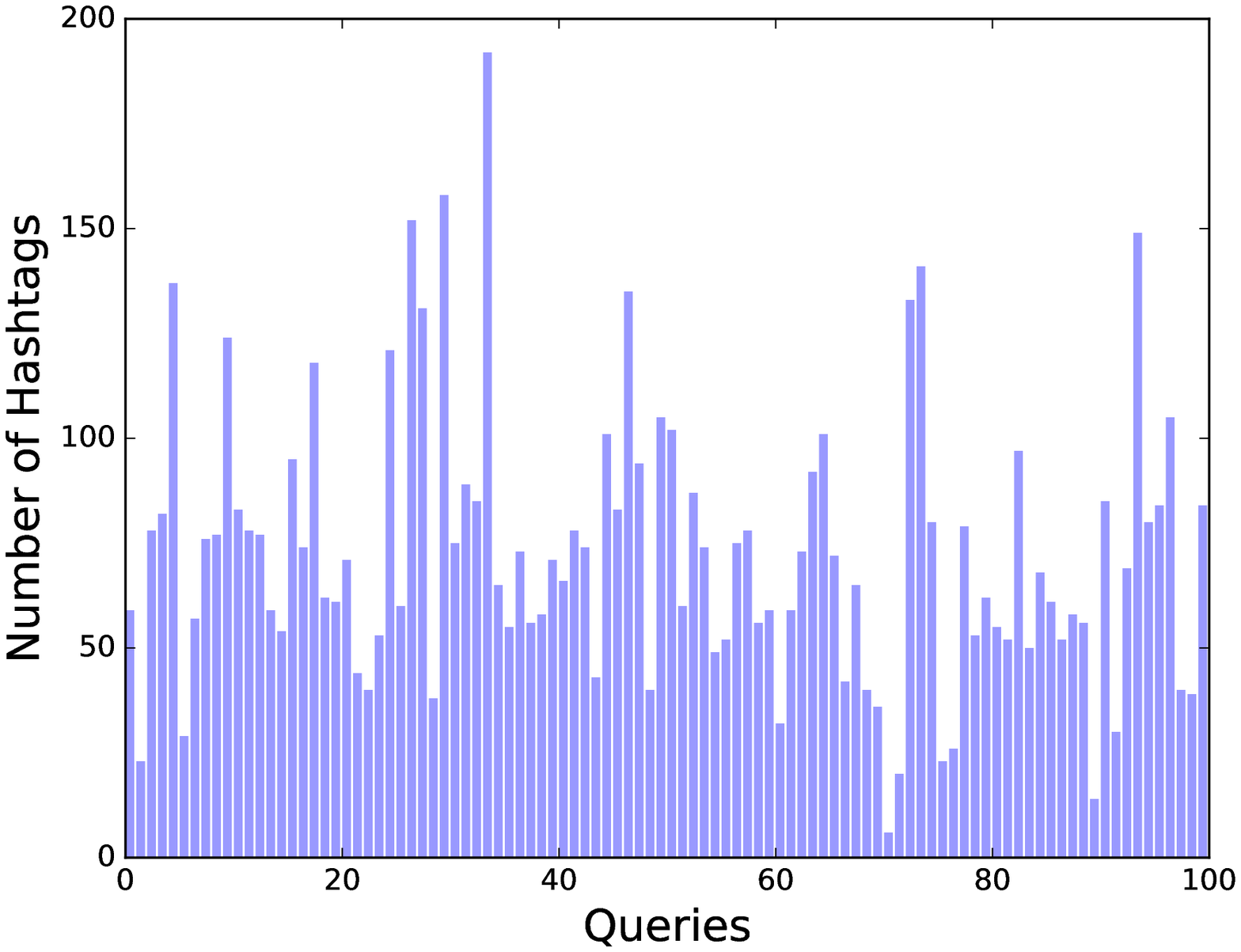}
\centerline{(a)Number of Hashtags}
\end{minipage}
\begin{minipage}[b]{0.49\linewidth}
\centering
\includegraphics[width=0.99\textwidth]{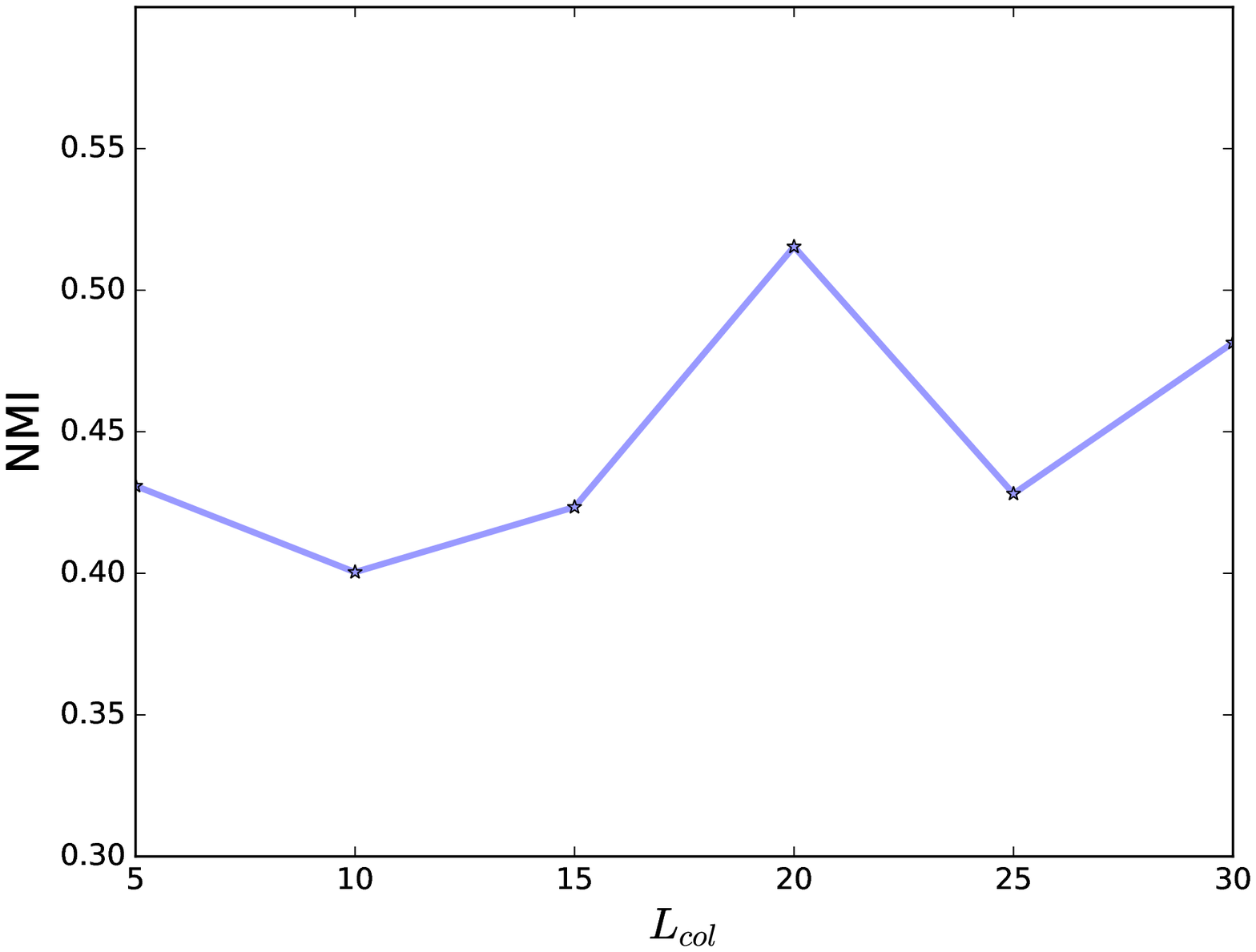}
\centerline{(b)Comparison on Different $L_{col}$}
\end{minipage}
\caption{Settings of Experiments on Hashtag-Topic Co-Clustering}
 \label{coclustering}
 \end{figure}
\begin{figure}[H]
\includegraphics[width=0.99\linewidth]{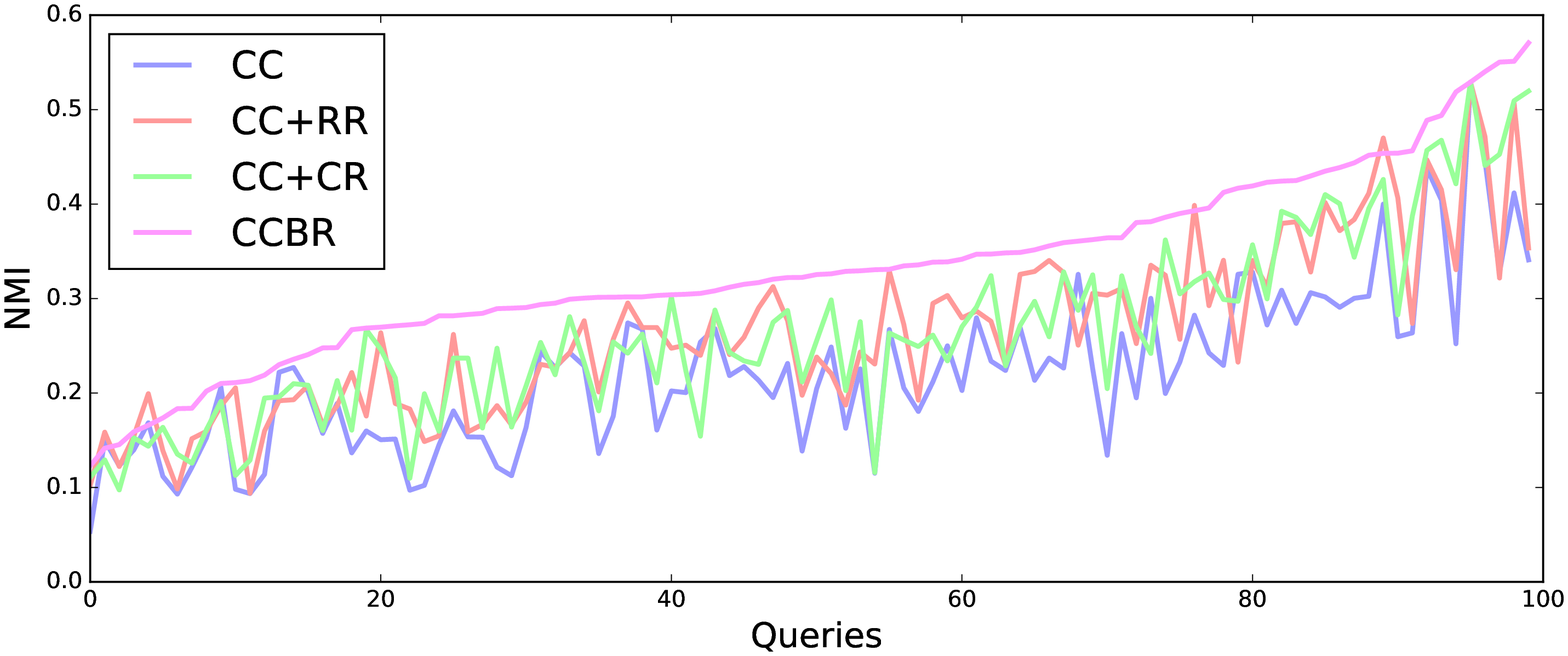}
\caption{Experiment Performance Comparision}
\label{expre}
\end{figure}

\begin{table}[H]
  \caption{Pearson correlation coefficient with different settings.}
  \label{P}
  \begin{tabular}{|c|c|c|c|c|}
   	\hline
     Method & CC & CC+CR & CC+RR  & CCBR\\
    \hline
    NMI \& \#hashtag & -0.689 & -0.633  & -0.643 & -0.526 \\
  \hline
\end{tabular}
\end{table}

\subsection{Results of Hashtag Clustering}
\subsubsection{Experimental Setting}
Given a hashtag-topic distribution $\textbf{H}$, the goal of Stage 2 is to generate hashtag clusters
$\{C_{1},C_{2},...,C_{L_{row}}\}$.
Therefore, we use Normalized Mutual Information (NMI) \cite{strehl2002cluster} as the evaluation metric.
Given a cluster label assignment $C_1$ on h objects, $H(C_1) = \sum_{i=1}^{|C_1|}P(i)\log(P(i))$ represents the entropy where $P(i) = |C_{1i}|/h$ denotes the probability that an object picked randomly from $C_1$ belongs to a cluster $C_{1i}$. Then the Normalized Mutual Information between two label assignments $C_1$ and $C_2$ is defined as:
\begin{equation}
\label{NMI}
NMI(C_1,C_2) = \frac{\sum_{i=1}^{|C_1|}\sum_{j=1}^{|C_2|}P(i,j)\log({\frac{P(i,j)}{P(i)P(j)}})}{\sqrt{H(C_1)H(C_2)}}
\end{equation}
where $H(C_2)= \sum_{j=1}^{|C_2|}P(j)\log(P(j))$, $P(j) = |{C_2}_{j}|/h$ and $P(i,j) = {|C_{1i}\cap C_{2j}|}/{h}$.

To make use of the NMI, the clusters of 100 randomly selected queries are manually labeled by 5 volunteers.
Regarding the labeling strategy, volunteers were under the guideline that they need to divide the hashtags into 5-9 clusters which determined by the number of hashtags of the search results as shown in Fig. \ref{coclustering}(a).
After that, we conducted the co-clustering process with the labeled truth as the number of hashtag clusters $L_{row}$.

Regarding the number of topic cluster $L_{col}$, we varied it from $5$ to $30$ with the step of 5 and reported the mean NMI in Fig. \ref{coclustering}(b). It is shown the clustering accuracy increases till $L_{col}=20$ and decreases afterwards. This indicates dividing
leaf topics into 20 clusters achieves best result and we thus select $L_{col} = 20$ in our experiments. For other parameters of co-clustering, we follow the empirical setting from \cite{banerjee2007generalized} and choose basis $\mathcal{C}_2$ and Squared Euclidean distance as $d_{\phi}$.

\subsubsection{Experimental Results and Analysis}
Performance comparison among different methods is shown in Fig.\ref{expre}. $CC$ is the original Bregman co-clustering method, $CC+RR$ is the co-clustering method with hashtag co-occurrence regularization,
$CC+CR$ is the original co-clustering method with intra-topic correlation regularization, and $CCBR $ represents the proposed method with bilateral regularization. The examined queries are shown in ascend order of the obtained NMI by $CCBR$. It is shown the curve $CCBR$ is above the other curves for most of the queries, demonstrating the advantages of bilateral regularization. Only considering intra-topic correlation or hashtag co-occurrence also improves the clustering performance over the Bregman co-clustering method to a certain extent.

Regarding the performance variance among different queries, we calculated the Pearson correlation coefficient between the obtained NMI and the number of hashtags for each query. The average results of different methods are shown in Table \ref{P}. The negative coefficients indicate that the query with larger number of hashtags are likely to achieve a lower NMI.

\subsection{Search Result Demonstration}
\subsubsection{Experimental Setting}
We focus on the evaluation of hashtag cluster ranking at this stage. After hashtag-topic clustering, the search result demonstration is executed on the hashtag clusters with the parameter $\psi = 0.5$. We utilize a widely used metric, Normalized Discounted Cumulative Gain (NDCG) for evaluation. NDCG is defined as:
\begin{equation}
\label{NDCG}
NDCG@k = \frac{1}{Z} \sum_{j=1}^k\frac{2^{r(j)}-1}{\log{(1+j)}}
\end{equation}
where $r(\cdot)$ is the relevance between the query and the ranked cluster which is calculated by Eqn. \ref{Qcost}.
\subsubsection{Experimental Results and Analysis}
\begin{figure}
\includegraphics[width=1\linewidth]{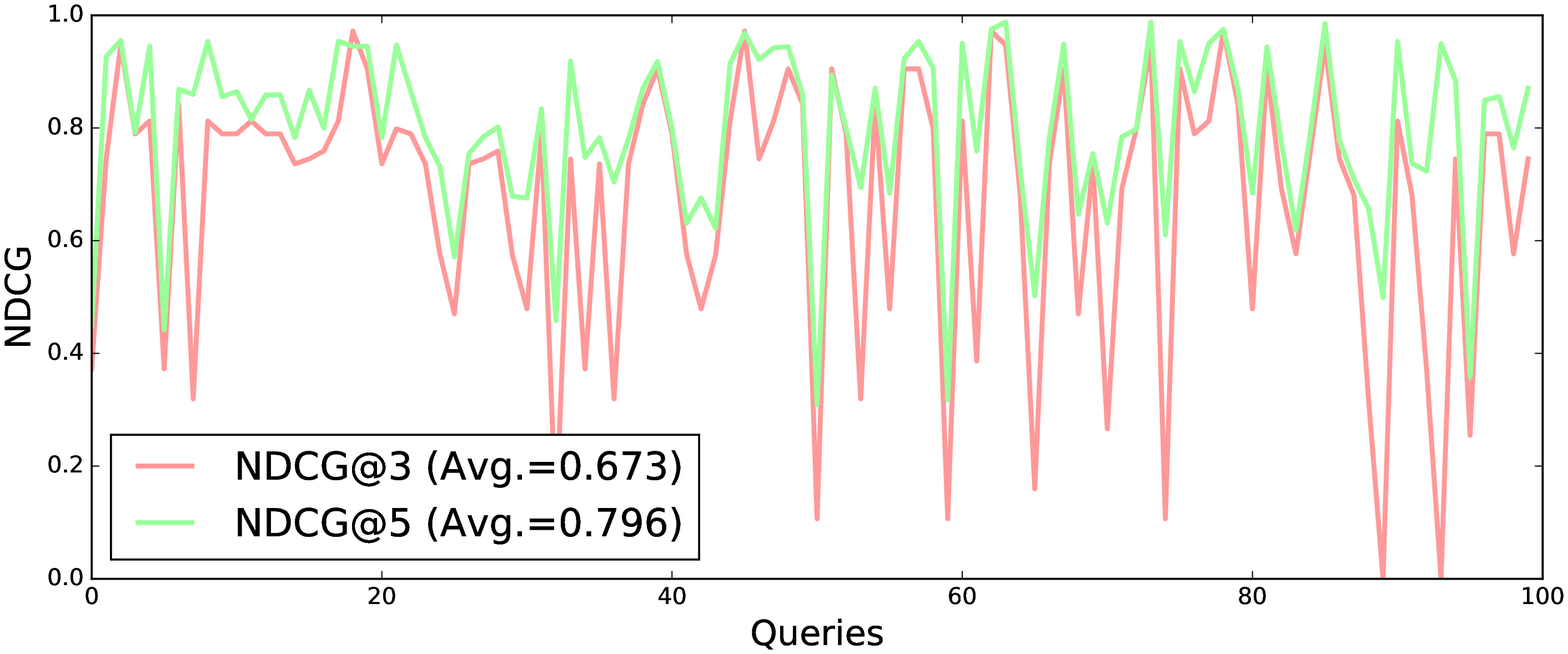}
\caption{NDCG for different queries}
\label{ndcg}
 \vspace{-3mm}
\end{figure}

To make use of NDCG, 5 volunteers were requested to vote for the top-5 appropriate clusters for each of the examined 100 queries. The ground-truth is averaged among the votes from volunteers. We show NDCG@3 and NDCG@5 in Fig. \ref{ndcg} for different queries. It can be observed that when examining top-5 clusters, the proposed rank method achieves a high average NDCG as $79.6\%$. Considering most queries have a ground-truth of 5-9 clusters, this high NDCG@5 indicates the practicality of the solution. When only examining the rank-1 cluster, the proposed rank method still achieves a satisfied performance with average NDCG@1=$37\%$. \footnote{Since NDCG@1 is either 0 or 1, we did not provide the detail results for each query in Fig.9}.

\begin{figure}
\begin{minipage}[b]{\linewidth}
\centering
\includegraphics[width=0.99\textwidth]{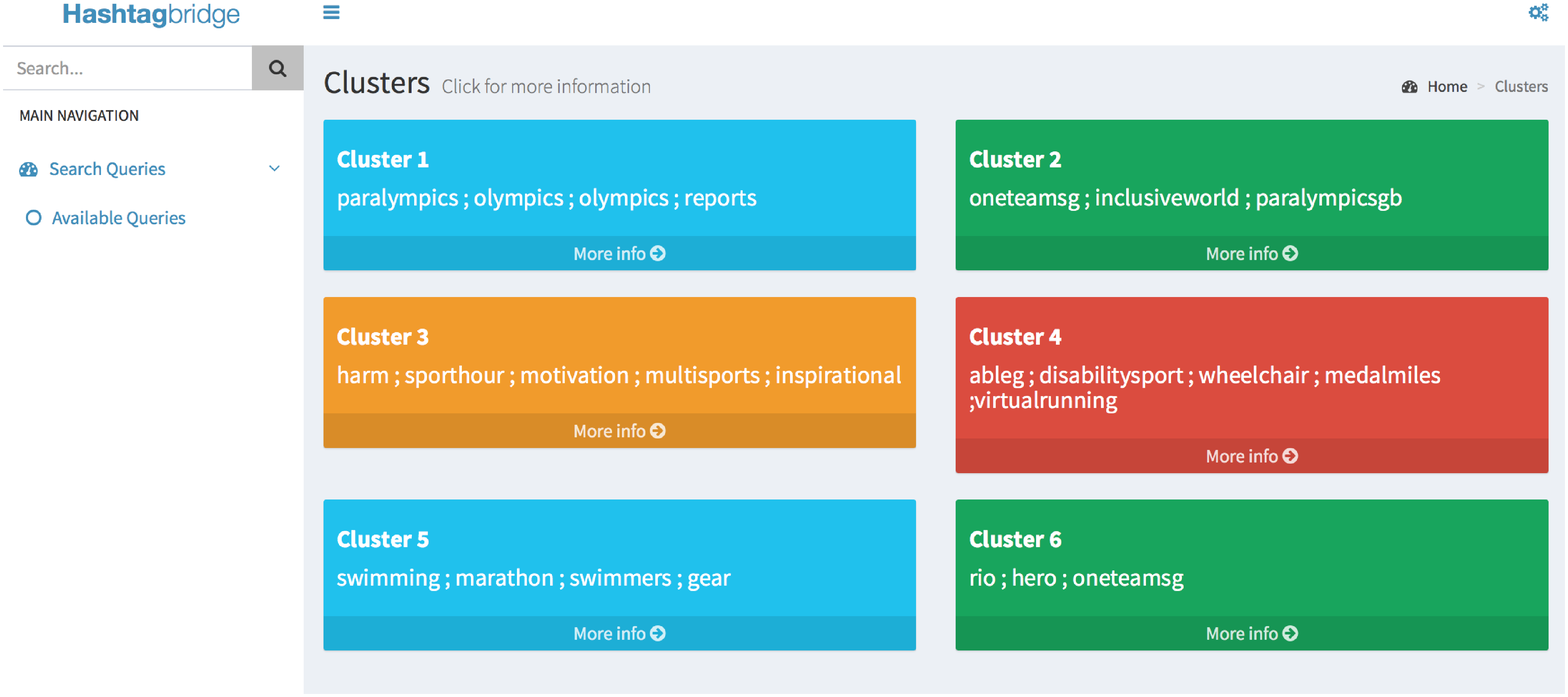}
\centerline{(a) Page for Clusters}
\end{minipage}
\begin{minipage}[b]{\linewidth}
\centering
\includegraphics[width=0.99\textwidth]{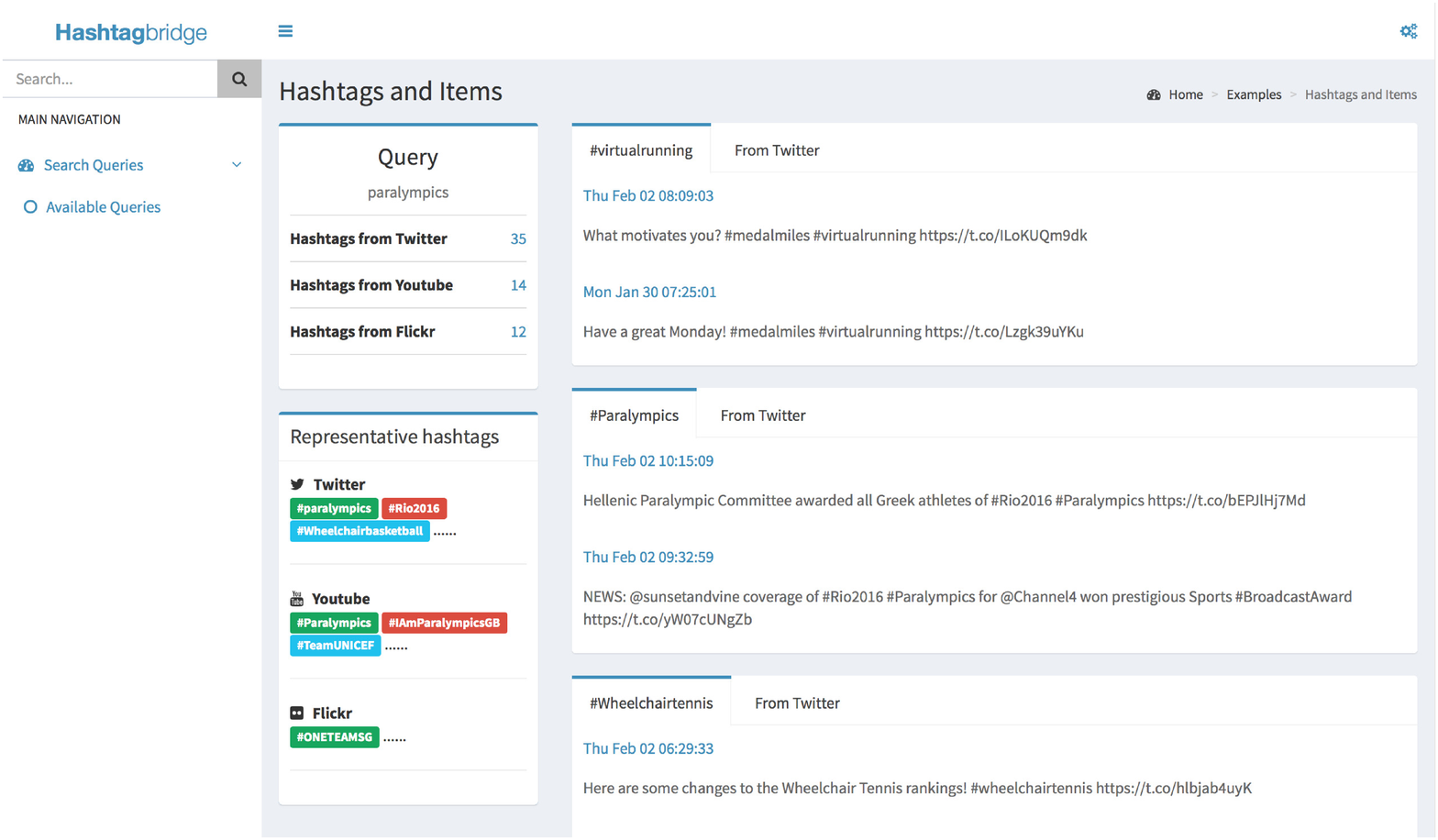}
\centerline{(b) Page for hashtags and items}
\end{minipage}
\vspace{-4mm}
\caption{An example of user interface}
 \label{demo}
 \vspace{-3mm}
 \end{figure}

More detailed results of the immersive search results are available at the web-based demo\footnote{\url{https://hashtagasbridge.github.io/Hashtag/}}. After issuing a event query, the related hashtags and information from Twitter, YouTube and Flickr are automatically collected and processed. Search results are organized following the cluster-hashtag-item hierachy, as illustrated in Fig. \ref{demo}. In Fig. \ref{demo}(a), the hashtag cluster is described by the words extracted according to Eqn.~\eqref{clusterword}. When clicking certain hashtag cluster, the assigned hashtags with related items within the cluster are displayed as in Fig. \ref{demo}(b).

\section{Conclusion and Future Work}
This study has positioned the problem of cross-OSN immersive search. A preliminary hashtag-centric solution is introduced. Hashtags are collected and exploited to organize the search results from different OSNs to help understand social events in a coarse-to-fine scheme. There is very long way to go before real-world application, and this work can be extended along several directions in the near future: (1) considering the time distribution of the collected hashtags, to visualize and track the evolution of events among OSNs; (2) exploring the social interaction potential of hashtag, e.g., analyzing the users who adopt the hashtags and creating event-oriented user channels to enrich the immersive search experience.

\bibliographystyle{ACM-Reference-Format}
\bibliography{sigconf-0420}

\end{document}